\documentclass{article}
\usepackage{amsmath,amsfonts,amssymb,color,dsfont,graphicx,makeidx,psfrag}
\newtheorem{conjecture}{Conjecture}
\newtheorem{definition}{Definition}
\newtheorem{example}{Example}
\newtheorem{exception}{Exception}
\newtheorem{theorem}{Theorem}
\newcommand{\ii}{\mathrm{i}}
\newcommand{\ij}{\mathrm{j}}

\DeclareMathOperator{\e}{e}

\begin{document}
\title{ Spectral resolution in hyperbolic orbifolds, quantum chaos,
  and cosmology} \author{ H.~Then \thanks{Institute of Physics,
    Carl-von-Ossietzky University, D-26111 Oldenburg, Germany}}
\date{}
\maketitle

\begin{abstract} We present a few subjects from physics that have one
  in common: the spectral resolution of the Laplacian. \end{abstract}

\section{Introduction}
If we look to nature, we observe dynamics and structure formation in
various aspects. Several theories exist that explain the growth of
structure quantitatively. On the smallest scale, it is quantum
mechanics that governs the dynamics, whereas on the largest scale the
evolution of our universe follows the Einstein field equations.

For all scales and almost any kind of observation we have equations at
hand that quantify our findings.  Many of these equations contain the
spectral resolution of the Laplacian. Just think of the Schr\"odinger
equation. In absence of forces or when the potential can be
transformed into the metric, the stationary Schr\"odinger equation
reduces to the eigenvalue equation of the Laplacian subject to some
boundary conditions.

Spectral resolution has important consequences on many topics.
Focusing on a few particular subjects, we start with an example from
thermodynamics and demonstrate the importance of the spectral density.
According to Weyl's law, the spectral density in leading order does
not depend on the shape of the boundaries.  In consequence, the
properties of thermodynamic systems are universal.

Less universal is the distribution of the eigenvalues. A central
subject of quantum chaos is to classify the distribution of
eigenvalues in dependence of whether the corresponding classical
system is chaotic.  In an explicit example, we compute the spectrum of
the Laplacian numerically and confirm a conjecture of arithmetic
quantum chaos.

Finally, we use the eigenvalues and eigenfunctions of the Laplacian to
compute the temperature fluctuations in the cosmic microwave
background.

\section{Thermodynamics}
The thermodynamic properties of an ideal gas can be deduced from the
logarithm of the partition function,
\begin{align*}
  \ln Z \equiv-\eta\sum_{\alpha}\ln(1-\eta
  z\mathrm{e}^{-\beta\varepsilon_{\alpha}}),
\end{align*}
where $\alpha$ stands for the quantum numbers of a single particle
with energy $\varepsilon_{\alpha}$, $z=\mathrm{e}^{\beta\mu}$ is the
fugacity, $\beta=\frac{1}{kT}$ is the inverse temperature, and $\eta$
is the statistic parameter that distinguishes between bosons,
$\eta=1$, and fermions, $\eta=-1$.

Introducing the spectral density
\begin{align*}
  d(\varepsilon):=\sum_{\alpha}\delta(\varepsilon-\varepsilon_{\alpha})
\end{align*}
the partition function can be expressed by an integral
\begin{align*}
  \ln Z=-\eta\int_{-\infty}^{\infty}d(\varepsilon) \ln(1-\eta
  z\mathrm{e}^{-\beta\varepsilon})\,\mathrm{d}\varepsilon
\end{align*}
with the advantage that it is easier to compute an integral than a sum
analytically.

\begin{example}[The ideal gas inside a cube]
  The quantum microstate of an ideal gas in thermal equilibrium is
  subject to the stationary Schr\"odinger equation
  \begin{align*} H\Psi=E\Psi \end{align*} with Dirichlet boundary
  conditions at $x,y,z=0$ and $x,y,z=L$.  Since the Hilbert space of
  an ideal gas separates, the Hamiltonian $H$ can be expressed by the
  sum of single-particle Hamiltonians
  $h_j=-\frac{\hbar^2}{2m}\Delta_j,\ j=1,\ldots,N$, where the
  Laplacian $\Delta_j$ acts only on the coordinates of the $j$-th
  particle.  Solving $h\psi=\varepsilon\psi$ yields the
  single-particle wave functions
  \begin{align*}
    \psi_{n_x,n_y,n_z}= \sqrt{\tfrac{2}{L}}\sin\Big(\frac{p_x
      x}{\hbar}\Big) \sqrt{\tfrac{2}{L}}\sin\Big(\frac{p_y
      y}{\hbar}\Big) \sqrt{\tfrac{2}{L}}\sin\Big(\frac{p_z
      z}{\hbar}\Big)
  \end{align*}
  and the single-particle eigenvalues
  \begin{align*}
    \varepsilon_{n_x,n_y,n_z}=\frac{p_x^2+p_y^2+p_z^2}{2m}
    =\frac{\hbar^2\pi^2}{2mL^2}(n_x^2+n_y^2+n_z^2)
  \end{align*}
  with the quantum numbers $n_x,n_y,n_z\in\mathds{N}$.
  
  In the thermodynamic limit, $V\to\infty$, the spectrum becomes
  continuous.  Consequently, the sum over the quantum numbers,
  \begin{align*}
    \sum_{\alpha}=\sum_{n_x}\sum_{n_y}\sum_{n_z}\sum_{m=-s}^{s},
  \end{align*}
  can be replaced by integrals
  \begin{align*}
    \sum_{n_x=1}^{\infty}\to\int_{0}^{\infty}\mathrm{d}n_x
    =\frac{L}{2\pi\hbar}\int_{-\infty}^{\infty}\mathrm{d}p_x\,,\qquad
    \mathrm{d}p_x=\frac{\hbar\pi}{L}\mathrm{d}n_x,
  \end{align*}
  resulting in
  \begin{align*}
    \sum_{\alpha}\to\frac{V}{(2\pi\hbar)^3}(2s+1)
    \int_{\mathds{R}^3}\mathrm{d}^3p,
  \end{align*}
  where $V=L^3$ is the volume and $s$ is the spin.
  
  Evaluating the spectral density
  \begin{align*}
    d(\varepsilon)=\sum_{\alpha}\delta(\varepsilon-\varepsilon_{\alpha})
    \to\frac{V}{(2\pi\hbar)^3}(2s+1)
    \int\mathrm{d}^3p\,\delta(\varepsilon-\varepsilon_{\alpha})
  \end{align*}
  with
  \begin{align*}
    \delta(\varepsilon-\varepsilon_{\alpha})=\delta(\varepsilon-\frac{p^2}{2m})
    =\frac{2m}{2|p|}\big(\delta(|p|-\sqrt{2m\varepsilon})
    +\delta(|p|+\sqrt{2m\varepsilon})\big).
  \end{align*}
  yields
  \begin{align*}
    d(\varepsilon)
    \sim(2s+1)\frac{(2m)^{\frac{3}{2}}}{4\pi^2\hbar^3}V\sqrt{\varepsilon}
    \quad \text{for} \quad \varepsilon\ge0 \qquad \text{and} \qquad
    d(\varepsilon)=0 \quad \text{for} \quad \varepsilon<0.
  \end{align*}
  Inserting it into the partition function gives
  \begin{align*}
    \ln Z=-\eta(2s+1)\frac{V}{\lambda^3}\frac{2}{\sqrt{\pi}}
    \int_{0}^{\infty}\mathrm{d}x\,\sqrt{x}\ln(1-\eta
    z\mathrm{e}^{-x})\,, \qquad
    \lambda:=\sqrt{\tfrac{2\pi\hbar^2}{m}\beta}\,.
  \end{align*}
  Taylor expanding the integrand
  \begin{align*}
    \ln(1-\eta z\mathrm{e}^{-x})=-\sum_{n=1}^{\infty} \frac{(\eta
      z\mathrm{e}^{-x})^n}{n}=\eta z\mathrm{e}^{-x}
    +\frac{1}{2}z^2\mathrm{e}^{-2x}+O(z^3),
  \end{align*}
  and evaluating the conditioning equation for the chemical potential,
  we have
  \begin{align*}
    N\equiv z\frac{\partial\ln Z}{\partial z}
    =(2s+1)\frac{V}{\lambda^3}z\big(1+\eta\frac{z}{2\sqrt{2}}+O(z^2)\big).
  \end{align*}
  Hence, the thermal equation of state
  \begin{align*}
    J:=-\frac{1}{\beta}\ln Z\equiv-pV,
  \end{align*}
  results in the virial expansion of the ideal gas
  \begin{align*}
    pV=NkT\big(1-\frac{\eta}{2s+1}\frac{N}{V}\frac{\lambda^3}{4\sqrt{2}}+\ldots
    \big).
  \end{align*}
  Interested in the inner energie, we obtain
  \begin{align*}
    U\equiv-\frac{\partial}{\partial\beta}\ln
    Z=\frac{3}{2}\frac{1}{\beta}\ln Z =\frac{3}{2}pV.
  \end{align*}
\end{example}
For further details and insight into thermodynamics, we refer the
reader to any standard textbook on statistical physics, e.g.\
\cite{then:LandauLifshitz1951}.

But what, if we ask for the ideal gas being inside a 3-sphere? Do we
need to compute the entire example again, beginning with the spectral
resolution of the Laplacian in a sphere?

\begin{theorem}[Weyl's law \cite{then:Weyl1912,then:Avakumovic1956}]
  If a quantum system is restricted to a finite volume $V$ in $D$
  dimensions, its level counting function
  \begin{align*}
    N(E):=\#\{\,E_i\,|\,E_i\le
    E\}\equiv\int_{-\infty}^{E}d(E)\,\mathrm{d}E
  \end{align*}
  is asymptotically equal to
  \begin{align*}
    N(E)\sim(2s+1)C_DVE^{\frac{D}{2}}
  \end{align*}
  in the semiclassical limit, $E\to\infty$, where
  \begin{align*}
    C_D=\frac{1}{\Gamma(\frac{D}{2}+1)}
    \big(\frac{m}{2\pi\hbar^2}\big)^{\frac{D}{2}}
  \end{align*}
  is a universal constant that only depends on the dimension $D$, but
  not on the specific shape of the boundary.
\end{theorem}

Weyl's law is nice. It tells us that the results of thermodynamics are
universal, i.e.\ they are independent of the boundary shape.

\section{Quantum chaos}
In terms of classical mechanics a system is described by specifying
the values of its coordinates and velocities, $x_1,\ldots,x_f$ and
$\dot{x}_1,\ldots,\dot{x}_f$, where $f$ is the number of degrees of
freedom.

If there exist $f$ linear independent constants of motion that are in
involution with each other, there is a set of canonical coordinates on
the phase space, the action-angle variables. The action variables are
constants of motion and the angle variables are the natural linear,
periodic coordinates on the torus. The motion on the torus is linear
in the angle variables, and the system is called {\em classically
  integrable}.

If there do not exist $f$ linear independent constants of motion, the
system is called {\em classical non-integrable} or {\em chaotic}. The
dynamics is non-linear and shows an exponential sensitivity to initial
conditions.

Being a limiting case of quantum mechanics, one might expect to see
the properties of classical mechanics in quantum theory. But it is not
this simple. The Schr\"odinger equation is linear and there is no
exponential sensitivity to initial conditions. Quantum chaos relies on
the behaviour of the corresponding classical system. A quantum system
is called {\em chaotic} if and only if the corresponding classical
system is non-integrable.

Central questions of quantum chaos concern the eigenvalue statistics
and the distribution of eigenvalues in the semiclassical limit.

We use the following assumptions: The quantum mechanical system is
desymmetrised with respect to all its unitary symmetries, and whenever
we examine the distribution of the eigenvalues, we regard them on the
scale of the mean level spacings.  Moreover, it is believed that after
desymmetrisation a generic quantum Hamiltonian possesses no degenerate
eigenvalues.
\begin{conjecture}[Berry, Tabor \cite{then:BerryTabor1976}]
  If the corresponding classical system is integrable, the eigenvalues
  behave like independent random variables and the distribution of the
  nearest-neighbour spacings is close to the Poisson distribution,
  i.e.  there is no level repulsion.
\end{conjecture}
\begin{conjecture}[Bohigas, Giannoni, Schmit
  \cite{then:BohigasGiannoniSchmit1984,then:BohigasGiannoniSchmit1986}]
  If the corresponding classical system is chaotic, the eigenvalues
  are distributed like the eigenvalues of hermitian random matrices.
  The corresponding ensembles depend only on the symmetries of the
  system:
  \begin{itemize}
  \item For chaotic systems without time-reversal invariance the
    distribution of the eigenvalues should be close to the
    distribution of the Gaussian Unitary Ensemble (GUE) which is
    characterised by a quadratic level repulsion.
  \item For chaotic systems with time-reversal invariance and integer
    spin the distribution of the eigenvalues should be close to the
    distribution of the Gaussian Orthogonal Ensemble (GOE) which is
    characterised by a linear level repulsion.
  \item For chaotic systems with time-reversal invariance and
    half-integer spin the distribution of the eigenvalues should be
    close to the distribution of the Gaussian Symplectic Ensemble
    (GSE) which is characterised by a quartic level repulsion.
  \end{itemize}
\end{conjecture}
These conjectures are very well confirmed by numerical calculations,
but several exceptions are known. Here are two examples:
\begin{exception}
  The harmonic oscillator is classically integrable, but its spectrum
  is equidistant.
\end{exception}
\begin{exception}
  The geodesic motion on surfaces with constant negative curvature
  provides a prime example for classical chaos. In some cases,
  however, the nearest-neighbour distribution of the eigenvalues of
  the Laplacian on these surfaces appears to be Poissonian.
\end{exception}
\begin{conjecture}[Arithmetic Quantum Chaos
  \cite{then:BogomolnyGeorgeotGiannoniSchmit1992,
    then:BolteSteilSteiner1992}]
  On surfaces of constant negative curvature that are generated by
  arithmetic fundamental groups, the distribution of the eigenvalues
  of the quantum Hamiltonian are close to the Poisson distribution.
  Due to level clustering small spacings occur comparably often.
\end{conjecture}

In the next sections, we compute numerically the eigenvalues and
eigenfunctions of the Laplacian that describe the quantum mechanics of
a point particle moving freely in the non-integrable three-dimensional
hyperbolic space of constant negative curvature generated by the
Picard group. The Picard group is arithmetic and we find that our
results are in accordance with the conjecture of arithmetic quantum
chaos.

\section{The modular surface}
For simplicity we first introduce the topology and geometry of the
two-dimensional surface of constant negative curvature that is
generated by the modular group \cite{then:Terras1985a}. It will then
be easy to carry over to the three-dimensional space of constant
negative curvature that is generated by the Picard group.

The construction begins with the upper half-plane,
\begin{align*} {\cal H}=\{(x,y)\in\mathds{R}^2, \quad y>0\},
\end{align*}
equipped with the hyperbolic metric of constant negative curvature
\begin{align*}
  ds^2=\frac{dx^2+dy^2}{y^2}.
\end{align*}

A free particle on the upper half-plane moves along geodesics, which
are straight lines and semicircles perpendicular to the $x$-axis,
respectively.  Expressing a point $(x,y)\in{\cal H}$ as a complex
number $z=x+\ii y$, all isometries of the hyperbolic metric are given
by the group of linear fractional transformations,
\begin{align*}
  z\mapsto\gamma z=\frac{az+b}{cz+d}, \quad a,b,c,d\in\mathds{R},
  \quad ad-bc=1,
\end{align*}
which is isomorphic to the group of matrices
\begin{align*}
  \gamma=\begin{pmatrix} a&b\\c&d \end{pmatrix}\in
  \operatorname{SL}(2,\mathds{R}),
\end{align*}
up to a common sign of the matrix entries,
\begin{align*}
  \operatorname{SL}(2,\mathds{R})/\{\pm1\}
  =\operatorname{PSL}(2,\mathds{R}).
\end{align*}

In analogy to the concept of a fundamental cell in a regular lattice
of a crystal we can introduce a fundamental domain of a discrete group
$\Gamma\subset\operatorname{PSL}(2,\mathds{R})$.
\begin{definition}
  A fundamental domain of the discrete group $\Gamma$ is an open
  subset ${\cal F}\subset{\cal H}$ with the following conditions: The
  closure of ${\cal F}$ meets each orbit $\Gamma z=\{\gamma z,\
  \gamma\in\Gamma\}$ at least once, ${\cal F}$ meets each orbit
  $\Gamma z$ at most once, and the boundary of ${\cal F}$ has Lebesgue
  measure zero.
\end{definition}
If we choose the group $\Gamma$ to be the modular group,
\begin{align*}
  \Gamma=\operatorname{PSL}(2,\mathds{Z}),
\end{align*}
which is generated by a translation and an inversion,
\begin{align*}
  \begin{pmatrix}1&1\\0&1\end{pmatrix}:\ z\mapsto z+1 \qquad
  \text{and} \qquad
  \begin{pmatrix}0&-1\\1&0\end{pmatrix}:\ z\mapsto-z^{-1},
\end{align*}
the fundamental domain of standard shape is
\begin{align*} {\cal F}=\{z=x+\ii y\in{\cal H}, \quad
  -\frac{1}{2}<x<\frac{1}{2}, \quad |z|>1\}.
\end{align*}

The isometric copies of the fundamental domain $\gamma{\cal F},\
\gamma\in\Gamma$, tessellate the upper half-plane completely without
any overlap or gap.  Identifying the fundamental domain ${\cal F}$ and
parts of its boundary with all its isometric copies $\gamma{\cal F},\
\forall\gamma\in\Gamma$, defines the topology to be the quotient space
$\Gamma\backslash{\cal H}$.  The quotient space $\Gamma\backslash{\cal
  H}$ can also be thought of as the fundamental domain ${\cal F}$ with
its faces glued according to the elements of the group $\Gamma$.

Any function being defined on the upper half-plane that is invariant
under linear fractional transformations,
\begin{align*}
  f(z)=f(\gamma z) \quad \forall\gamma\in\Gamma,
\end{align*}
can be identified with a function living on the quotient space
$\Gamma\backslash{\cal H}$ and vice versa.

With the hyperbolic metric the quotient space $\Gamma\backslash{\cal
  H}$ inherits the structure of an orbifold. An orbifold locally looks
like a manifold, with the exception that it is allowed to have
elliptic fixed-points.

The orbifold of the modular group has one parabolic and two elliptic
fixed-points,
\begin{align*}
  z=\ii\infty, \quad z=\ii, \quad \text{and} \quad
  z=\frac{1}{2}+\ii\frac{\sqrt{3}}{2}.
\end{align*}
The parabolic one fixes a cusp at $z=\ii\infty$ which is invariant
under the parabolic element
\begin{align*}
  \begin{pmatrix}1&1\\0&1\end{pmatrix}.
\end{align*}
Hence, the orbifold of the modular group is non-compact.  The volume
element corresponding to the hyperbolic metric reads
\begin{align*}
  d\mu=\frac{dx dy}{y^2},
\end{align*}
such that the volume of the orbifold $\Gamma\backslash{\cal H}$ is
finite,
\begin{align*}
  \operatorname{vol}(\Gamma\backslash{\cal H})=\frac{\pi}{3}.
\end{align*}

Scaling the units such that $\hbar=1$ and $2m=1$, the stationary
Schr\"{o}dinger equation which describes the quantum mechanics of a
point particle moving freely in the orbifold $\Gamma\backslash{\cal
  H}$ becomes
\begin{align*}
  (\Delta+\lambda)f(z)=0,
\end{align*}
where the hyperbolic Laplacian is given by
\begin{align*}
  \Delta=y^2(\frac{\partial^2}{\partial x^2}
  +\frac{\partial^2}{\partial y^2})
\end{align*}
and $\lambda$ is the scaled energy.  We can relate the eigenvalue
problem defined on the orbifold $\Gamma\backslash{\cal H}$ to the
eigenvalue problem defined on the upper-half space, with the
eigenfunctions being subject to the automorphy condition relative to
the discrete group $\Gamma$,
\begin{align*}
  f(\gamma z)=f(z) \quad \forall\gamma\in\Gamma.
\end{align*}
In order to avoid solutions that grow exponentially in the cusp, we
impose the boundary condition
\begin{align*}
  f(z)=O(y^\kappa) \quad \text{for} \quad z\to\ii\infty,
\end{align*}
where $\kappa$ is some positive constant.

The solutions of this eigenvalue problem can be identified with Maass
waveforms \cite{then:Maass1949a}. Numerically, they are most
efficiently computed using Hejhal's algorithm \cite{then:Hejhal1999}.

\section{The Picard surface} \label{then-sec:5} In the
three-dimensional case one considers the upper-half space,
\begin{align*} {\cal H}=\{(x_0,x_1,y)\in\mathds{R}^3, \quad y>0\}
\end{align*}
equipped with the hyperbolic metric
\begin{align*}
  ds^2=\frac{dx_0^2+dx_1^2+dy^2}{y^2}.
\end{align*}
The geodesics of a particle moving freely in the upper half-space are
straight lines and semicircles perpendicular to the $x_0$-$x_1$-plane,
respectively.

Expressing any point $(x_0,x_1,y)\in{\cal H}$ as a quaternion,
$z=x_0+\ii x_1+\ij y$, with the multiplication defined by $\ii^2=-1,\
\ij^2=-1,\ \ii\ij+\ij\ii=0$, all motions in the upper half-space are
given by linear fractional transformations
\begin{align*}
  z\mapsto\gamma z=(az+b)(cz+d)^{-1}, \quad a,b,c,d\in\mathds{C},
  \quad ad-bc=1.
\end{align*}
The group of these transformations is isomorphic to the group of
matrices
\begin{align*}
  \gamma=\begin{pmatrix}a&b\\c&d\end{pmatrix}\in
  \operatorname{SL}(2,\mathds{C})
\end{align*}
up to a common sign of the matrix entries,
\begin{align*}
  \operatorname{SL}(2,\mathds{C})/\{\pm1\}
  =\operatorname{PSL}(2,\mathds{C}).
\end{align*}
The motions provided by the elements of
$\operatorname{PSL}(2,\mathds{C})$ exhaust all orientation preserving
isometries of the hyperbolic metric on ${\cal H}$.

We now choose the discrete group
$\Gamma\subset\operatorname{PSL}(2,\mathds{C})$ that is generated by
the cosets of the following elements,
\begin{align*}
  \begin{pmatrix}1&1\\0&1\end{pmatrix}, \quad
  \begin{pmatrix}1&\ii\\0&1\end{pmatrix}, \quad
  \begin{pmatrix}0&-1\\1&0\end{pmatrix},
\end{align*}
which yield two translations and one inversion,
\begin{align*}
  z\mapsto z+1, \quad z\mapsto z+\ii, \quad z\mapsto-z^{-1}.
\end{align*}
This group $\Gamma$ is called the Picard group. The three motions
generating $\Gamma$, together with the coset of the element
\begin{align*}
  \begin{pmatrix}\ii&0\\0&-\ii\end{pmatrix}
\end{align*}
that is isomorphic to the symmetry
\begin{align*}
  z=x+\ij y\mapsto\ii z\ii=-x+\ij y,
\end{align*}
can be used to construct the fundamental domain of standard shape
\begin{align*} {\cal F}=\{z=x_0+\ii x_1+\ij y\in{\cal H}, \quad
  -\frac{1}{2}<x_0<\frac{1}{2}, \quad 0<x_1<\frac{1}{2}, \quad
  |z|>1\}.
\end{align*}

Identifying the faces of the fundamental domain according to the
elements of the group $\Gamma$ leads to a realisation of the quotient
space $\Gamma\backslash{\cal H}$.

With the hyperbolic metric the quotient space $\Gamma\backslash{\cal
  H}$ inherits the structure of an orbifold that has one parabolic and
four elliptic fixed-points,
\begin{align*}
  z=\ij\infty, \quad z=\ij, \quad z=\frac{1}{2}+\ij\sqrt{\frac{3}{4}},
  \quad z=\frac{1}{2}+\ii\frac{1}{2}+\ij\sqrt{\frac{1}{2}}, \quad
  z=\ii\frac{1}{2}+\ij\sqrt{\frac{3}{4}}.
\end{align*}
The parabolic fixed-point corresponds to a cusp at $z=\ij\infty$ that
is invariant under the parabolic elements
\begin{align*}
  \begin{pmatrix}1&1\\0&1\end{pmatrix} \quad \text{and} \quad
  \begin{pmatrix}1&\ii\\0&1\end{pmatrix}.
\end{align*}
The volume element deriving from the hyperbolic metric reads
\begin{align*}
  d\mu=\frac{dx_0dx_1dy}{y^3},
\end{align*}
such that the volume of the non-compact orbifold
$\Gamma\backslash{\cal H}$ is finite,
\begin{align*}
  \operatorname{vol}(\Gamma\backslash{\cal
    H})=\frac{\zeta_K(2)}{4\pi^2} \simeq0.3053218647...,
\end{align*}
where
\begin{align*}
  \zeta_K(s)=\frac{1}{4}\sum_{\nu\in\mathds{Z}[\ii]-\{0\}}
  (\nu\bar{\nu})^{-s}, \quad \Re s>1,
\end{align*}
is the Dedekind zeta function.

We are interested in the square-integrable eigenfunctions of the
Laplacian,
\begin{align*}
  \Delta=y^2\big(\frac{\partial^2}{\partial x_0^2}+\frac{\partial^2}
  {\partial x_1^2}+\frac{\partial^2}{\partial
    y^2}\big)-y\frac{\partial} {\partial y},
\end{align*}
that determine the quantum mechanics of a point particle moving
freely in the orbi\-fold $\Gamma\backslash{\cal H}$. We identify the
solutions with Maass waveforms \cite{then:Maass1949b}.

Since Maass waveforms are automorphic, and therefore periodic in $x_0$
and $x_1$, it follows that they can be expanded into a Fourier series,
\begin{align*}
  f(z)=u(y)+\sum_{\beta\in\mathds{Z}[\ii]-\{0\}} a_{\beta}yK_{\ii
    r}(2\pi|\beta|y)\e^{2\pi\ii\Re\beta x},
\end{align*}
where
\begin{align*}
  u(y)=\begin{cases} b_0 y^{1+\ii r}+b_1 y^{1-\ii r}& \text{if
      $r\not=0$},\\ b_2 y+b_3 y\ln y& \text{if $r=0$}.
  \end{cases}
\end{align*}
$K_{\ii r}(x)$ is the K-Bessel function whose order is connected with
the eigenvalue $\lambda$ by
\begin{align*}
  \lambda=r^2+1.
\end{align*}
If a Maass waveform vanishes in the cusp,
\begin{align*}
  \lim_{z\to\ij\infty}f(z)=0,
\end{align*}
it is called a Maass cusp form. Maass cusp forms are square integrable
over the fundamental domain, $\langle f,f\rangle<\infty$, where
\begin{align*}
  \langle f,g\rangle=\int_{\Gamma\backslash{\cal H}}\bar{f}g\,d\mu
\end{align*}
is the Petersson scalar product.

According to the Roelcke-Selberg spectral resolution of the Laplacian
\cite{then:Roelcke1966-1967}, its spectrum contains both a discrete
and a continuous part. The discrete part is spanned by the constant
eigenfunction $f_0$ and a countable number of Maass cusp forms
$f_1,f_2,f_3,\ldots$ which we take to be ordered with increasing
eigenvalues, $0=\lambda_0<\lambda_1\le\lambda_2\le\lambda_3\le\ldots$.
The continuous part of the spectrum $\lambda\ge1$ is spanned by the
Eisenstein series $E(z,1+\ii r)$ which are known analytically
\cite{then:ElstrodtGrunewaldMennicke1985}.  The Fourier coefficients
of the functions $\Lambda_K(1+\ii r)E(z,1+\ii r)$ are given by
\begin{align*}
  b_0=\Lambda_K(1+\ii r), \quad b_1=\Lambda_K(1-\ii r), \quad
  a_{\beta}=2\sum_{\substack{\lambda,\mu\in\mathds{Z}[\ii] \\
      \lambda\mu=\beta}} \big|\frac{\lambda}{\mu}\big|^{\ii r},
\end{align*}
where
\begin{align*}
  \Lambda_K(s)=4\pi^{-s}\Gamma(s)\zeta_K(s)
\end{align*}
has an analytic continuation into the complex plane except for a pole
at $s=1$.

Normalising the Maass cusp forms according to
\begin{align*}
  \langle f_n,f_n\rangle=1,
\end{align*}
we can expand any square integrable function $\phi\in
L^2(\Gamma\backslash{\cal H})$ in terms of Maass waveforms,
\begin{align*}
  \phi(z)=\sum_{n\ge0}\langle f_n,\phi\rangle f_n(z)+\frac{1}{2\pi\ii}
  \int_{\Re s=1}\langle E(\cdot,s),\phi\rangle E(z,s)\,ds.
\end{align*}

The eigenvalues and their associated Maass cusp forms are not known
analytically. Thus, one has to approximate them numerically. By making
use of the Hecke operators and the multiplicative relations among the
coefficients, Steil \cite{then:Steil1999} obtained a non-linear system
of equations which allowed him to compute $2545$ consecutive
eigenvalues. In \cite{then:Then2006}, we extend these computations
with the use of a variant of Hejhal's algorithm
\cite{then:Hejhal1999}.  Our favourite explanation of the algorithm is
published in \cite{then:Then2005}.

\section{Results} \label{then-sec:6} The modular surface, i.e.\ the
two-dimensional hyperbolic orbifold that is generated by the modular
group, has a reflection symmetry.  This reflection symmetry commutes
with the Laplacian. Consequently, the eigenfunctions fall into two
symmetry classes. We call an eigenfunction to be even or odd depending
on whether $f(-x+\ii y)=f(x+\ii y)$ or $f(-x+\ii y)=-f(x+\ii y)$
holds. We also call the corresponding eigenvalue to be even or odd,
respectively.

In table \ref{then-tab:1}, the first ten consecutive even and the
first ten consecutive odd eigenvalues of the Laplacian on the modular
surface are listed.
\begin{table} \caption{The first ten even and the first ten odd
    eigenvalues of the negative Laplacian on the modular surface.
    Listed is $r$, related to the eigenvalue via
    $\lambda=r^2+\frac{1}{4}$.} \label{then-tab:1}
  \begin{scriptsize} \begin{align*} &\text{even} & &\text{odd} \\ \\
      13&.77975135189 &  9&.53369526135 \\
      17&.73856338106 & 12&.17300832468 \\
      19&.42348147083 & 14&.35850951826 \\
      21&.31579594020 & 16&.13807317152 \\
      22&.78590849419 & 16&.64425920190 \\
      24&.11235272984 & 18&.18091783453 \\
      25&.82624371271 & 19&.48471385474 \\
      26&.15208544922 & 20&.10669468255 \\
      27&.33270808315 & 21&.47905754475 \\
      28&.53074769292 & 22&.19467397757 \\
    \end{align*} \end{scriptsize} \end{table}
\begin{table} \caption{The first few eigenvalues of the negative
    Laplacian on the Picard surface. Listed is $r$, related to the
    eigenvalues via $\lambda=r^2+1$.}
  \label{then-tab:2}
  \begin{scriptsize} \begin{align*}
      {\mathbf D}& & {\mathbf G}& & {\mathbf C}& & {\mathbf H}& \\ \\
      \ 8&.55525104 & & & \ 6&.62211934 \\
      11&.10856737 & & & 10&.18079978 \\
      12&.86991062 & & & 12&.11527484 & 12&.11527484 \\
      14&.07966049 & & & 12&.87936900 \\
      15&.34827764 & & & 14&.14833073 \\
      15&.89184204 & & & 14&.95244267 & 14&.95244267 \\
      17&.33640443 & & & 16&.20759420 \\
      17&.45131992 & 17&.45131992 & 16&.99496892 & 16&.99496892 \\
      17&.77664065 & & & 17&.86305643 & 17&.86305643 \\
      19&.06739052 & & & 18&.24391070 \\
      19&.22290266 & & & 18&.83298996 \\
      19&.41119126 & & & 19&.43054310 & 19&.43054310 \\
      20&.00754583 & & & 20&.30030720 & 20&.30030720 \\
      20&.70798880 & 20&.70798880 & 20&.60686743 \\
      20&.81526852 & & & 21&.37966055 & 21&.37966055 \\
      21&.42887079 & & & 21&.44245892 \\
      22&.12230276 & & & 21&.83248972 & 21&.83248972 \\
      22&.63055256 & & & 22&.58475297 & 22&.58475297 \\
      22&.96230105 & 22&.96230105 & 22&.85429195 \\
      23&.49617692 & & & 23&.49768305 & 23&.49768305 \\
    \end{align*} \end{scriptsize} \end{table}

The Picard surface, i.e.\ the three-dimensional hyperbolic orbifold
that is generated by the Picard group, has two symmetries such that
the eigenfunctions of the Laplacian fall into four symmetry classes.
The eigenfunctions and their corresponding eigenvalues are called to
be of symmetry class ${\mathbf D}$, ${\mathbf G}$, ${\mathbf C}$, or
${\mathbf H}$ depending on whether
\begin{align*}
  &f(x+\ij y)=f(\ii x+\ij)=f(-\bar{x}+\ij y), \\
  &f(x+\ij y)=f(\ii x+\ij)=-f(-\bar{x}+\ij y), \\
  &f(x+\ij y)=-f(\ii x+\ij)=f(-\bar{x}+\ij y), \\
  &f(x+\ij y)=-f(\ii x+\ij)=-f(-\bar{x}+\ij y)
\end{align*} holds, respectively.

The first few consecutive eigenvalues of each symmetry class of the
Laplacian on the Picard surface are listed in table \ref{then-tab:2}.

Examining the eigenvalues, we find that there occur degenerate
eigenvalues. The smallest degenerate eigenvalue of the Laplacian on
the Picard surface is $\lambda=12.11527484^2+1$. The degeneracies
result from the symmetries of the Picard surface. If desymmetrised,
i.e.\ within a given symmetry class, there do not exist any degenerate
eigenvalues. An explanation is given in \cite{then:Then2006}.

Consider the level counting function
\begin{align*} N(r):=\#\{\,r_i\ |\ r_i\le r\} \end{align*} and split
it into two parts
\begin{align*} N(r)=\bar{N}(r)+N_{fluc}(r). \end{align*} Here
$\bar{N}$ is a smooth function describing the average increase in the
number of levels and $N_{fluc}$ describes the fluctuations around the
mean such that
\begin{align*}
  \lim_{R\to\infty}\frac{1}{R}\int_{1}^{R}N_{fluc}(r)dr=0.
\end{align*}
According to Weyl's law and higher order corrections for the Picard
surface found by Matthies \cite{then:Matthies1995}, the average
increase in the number of levels is given by
\begin{align*}
  \bar{N}(r)=\tfrac{\operatorname{vol}({\cal F})}{6\pi^2}r^3 +a_2 r
  \log r+a_3 r+a_4
\end{align*}
with the constants
\begin{align*}
  a_2&=-\tfrac{3}{2\pi}, \\
  a_3&=\tfrac{1}{\pi}[\tfrac{13}{16}\log 2
  +\tfrac{7}{4}\log\pi-\log\Gamma(\tfrac{1}{4})
  +\tfrac{2}{9}\log(2+\sqrt{3})+\tfrac{3}{2}], \\
  a_4&=-\tfrac{1}{2}.
\end{align*}

If desymmetrised into the four symmetry classes, the average increase
in the number of levels is given by
\begin{align*}
  \bar{N}(r)=\tfrac{1}{4}\tfrac{\operatorname{vol}({\cal
      F})}{6\pi^2}r^3 +b_1 r^2+b_2 r \log r+b_3 r+b_4
\end{align*}
with the constants depending on the symmetry class as listed in table
\ref{then-tab:3}.
\begin{table} \caption{The constants for the higher order corrections
    to Weyl's law for each of the four symmetry classes of the Picard
    surface.  The constants $b_1$ and $b_2$ are known analytically
    \cite{then:Matthies1995}, whereas the constants $b_3$ and $b_4$
    have been approximated numerically \cite{then:Then2006}.}
  \label{then-tab:3}
  \begin{align*} & & b_1 & & b_2 & & b_3\ \qquad & & b_4\qquad \\ \\
    {\mathbf D}& &
    \tfrac{1}{24} & & -\tfrac{13}{8\pi} & & 0.8639... & & -0.288... \\
    {\mathbf G}& &
    -\tfrac{1}{24} & & \tfrac{3}{8\pi} & & 0.0285... & & -0.184... \\
    {\mathbf C}& &
    \tfrac{1}{96} & & -\tfrac{1}{8\pi} & & 0.0150... & & -0.062... \\
    {\mathbf H}& &
    -\tfrac{1}{96} & & -\tfrac{1}{8\pi} & & 0.0702... & & 0.034... \\
  \end{align*} \end{table}
\begin{figure}
  \includegraphics[height=5cm,angle=-90]{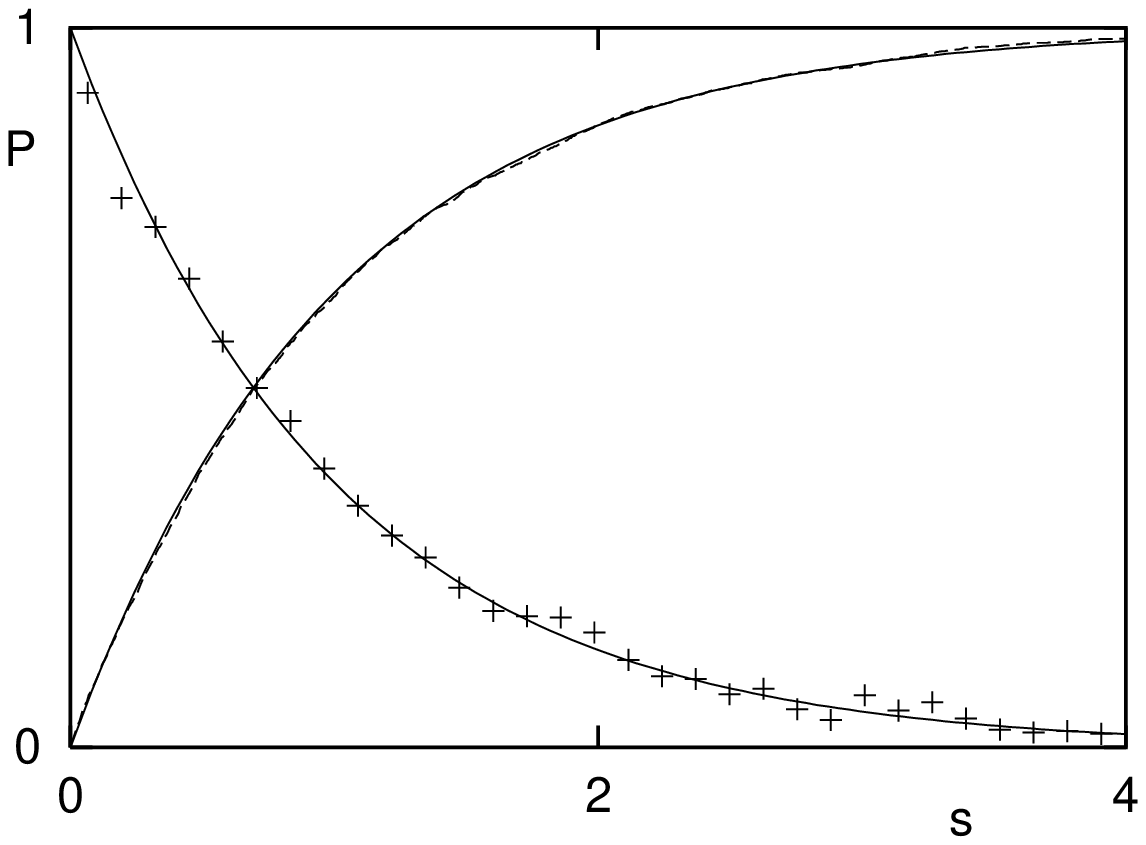}
  \qquad
  \includegraphics[height=5cm,angle=-90]{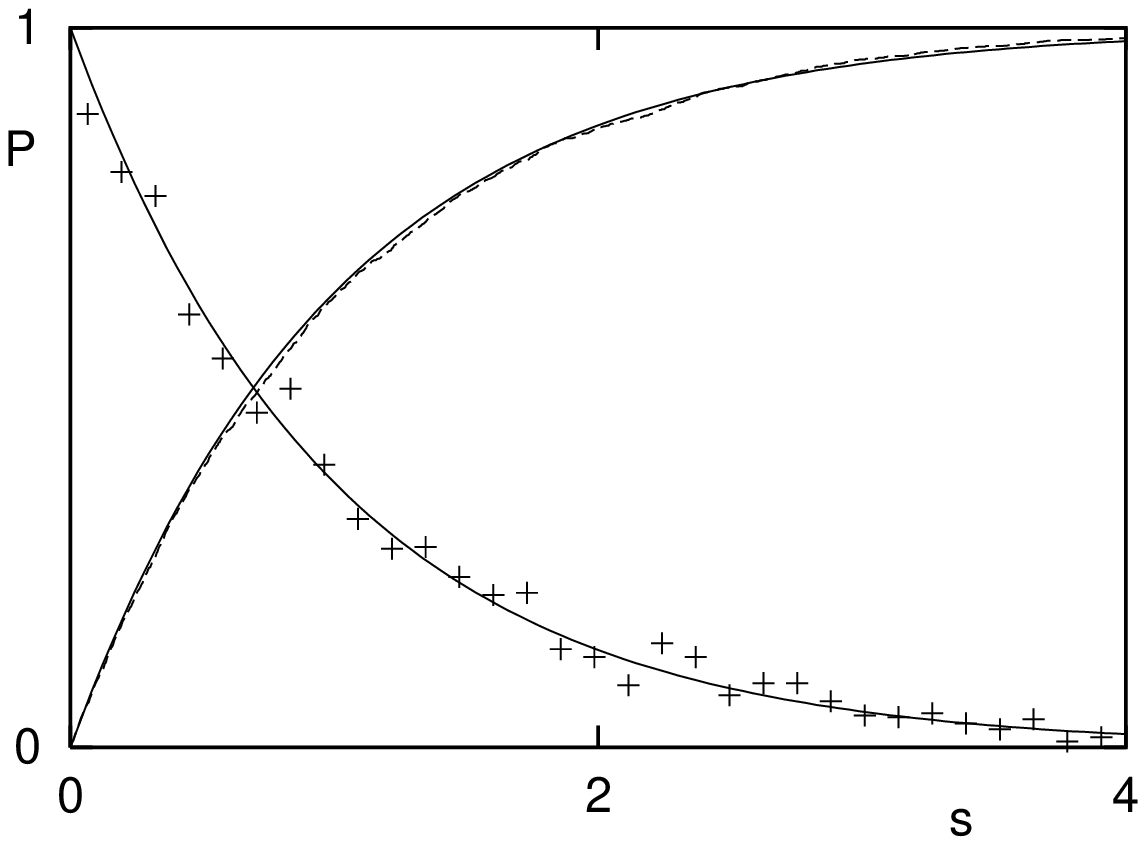}
  \\
  \includegraphics[height=5cm,angle=-90]{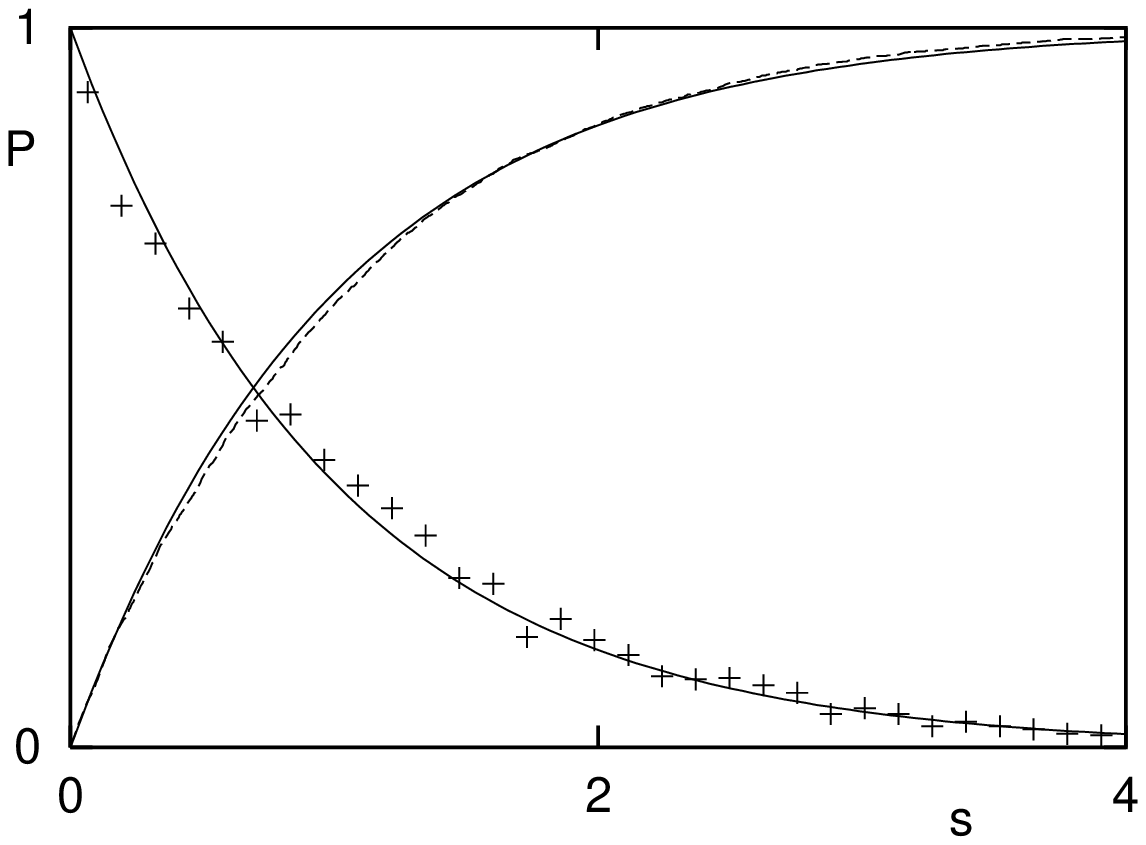}
  \qquad
  \includegraphics[height=5cm,angle=-90]{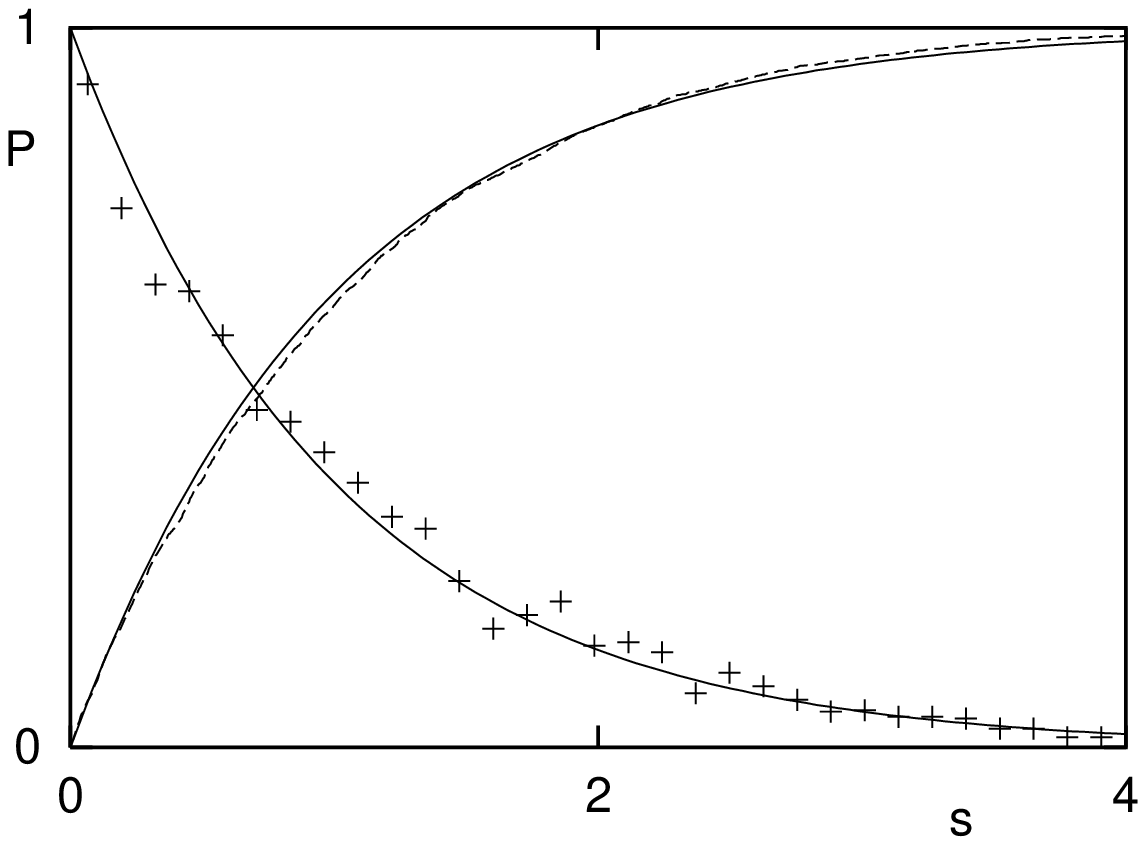}
  \caption{Level spacing distributions for the symmetry classes
    ${\mathbf D}$ (top left), ${\mathbf G}$ (top right), ${\mathbf C}$
    (bottom left), and ${\mathbf H}$ (bottom right) of the Picard
    surface. The abscissae display the spacings $s$. The small crosses
    are the histograms of altogether $13949$ nearest-neighbour
    spacings. The dashed curves starting at the origins are the
    integrated distributions. For comparison, the full curves show the
    Poisson distribution.} \label{then-fig:1}
\end{figure}

Unfolding the spectrum,
\begin{align*}
  x_i:=\bar{N}(r_i),
\end{align*}
we are able to examine the distribution of the eigenvalues on the
scale of the mean level spacings. Defining the sequence of
nearest-neighbour level spacings with mean value $1$ as $i\to\infty$,
\begin{align*}
  s_i:=x_{i+1}-x_i,
\end{align*}
we find that the spacing distribution comes close to that of a Poisson
random process,
\begin{align*}
  P_{\text{Poisson}}(s)=\mathrm{e}^{-s},
\end{align*}
see figure \ref{then-fig:1}, in accordance with the conjecture of
arithmetic quantum chaos.

\section{Cosmology}
In the remaining sections we apply the eigenvalues and eigenfunctions
of the Laplacian to a perturbed Ro\-bert\-son-Wal\-ker universe and
compute the temperature fluctuations in the cosmic microwave
background (CMB).
\begin{figure} \centering \includegraphics[height=10cm]{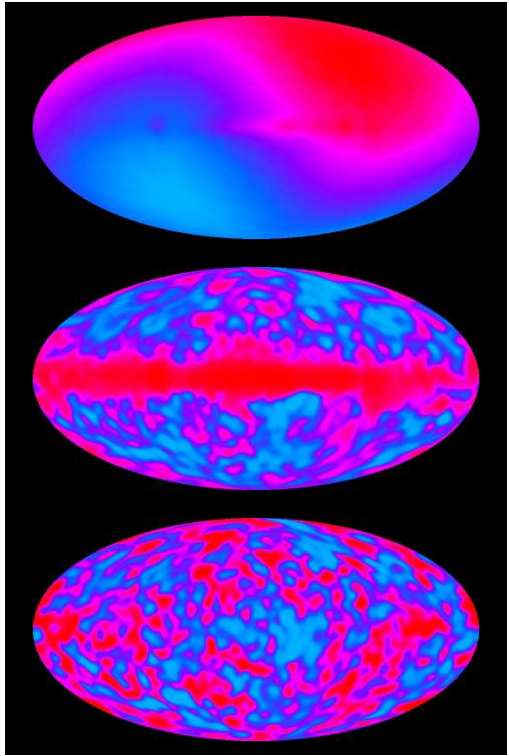}
  \caption{Sky maps of the temperature fluctuations in the CMB as
    observed by the NASA satellite mission COBE. The sky map on the
    top shows the dipole anisotropy after the mean background
    temperature of $T_0=2.725\,\text{K}$ has been subtracted. The
    amplitude of the dipole anisotropy is about $3\,\text{mK}$. Also
    subtracting the dipole yields the sky map in the middle. One sees
    the small temperature fluctuations whose amplitude is roughly
    $30\,\mu\text{K}$. But one also sees a lot of foreground
    contamination along the equator that comes from nearby stars in
    our galaxy. After removing the foregrounds one finally gets the
    sky map on the bottom showing the temperature fluctuations in the
    CMB. Downloaded from \cite{then:LAMBDA2003}.} \label{then-fig:2}
\end{figure}
\begin{figure} \centering \input{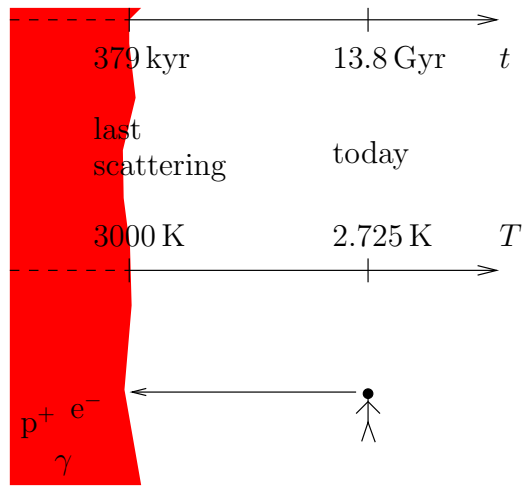} \caption{The
    expanding universe. At the time of last scattering occured a phase
    transition from an opaque to a transparent universe.}
  \label{then-fig:3} \end{figure} \par
The CMB is a relic from the primeval fireball of the early universe.
It is the light that comes from the time when the universe was
$379\,000$ years old. It was predicted by Gamow in 1948 and explained
in detail by Peebles \cite{then:Peebles1965}. In 1978, Penzias and
Wilson won the Nobel Prize of Physics for first measuring the CMB at a
wavelength of $7.35\,\text{cm}$. Within the resolution of their
experiment they found the CMB to be completely isotropic over the
whole sky. Later with the much better resolution of the NASA satellite
mission Cosmic Background Explorer (COBE), Smoot et
al.~\cite{then:SmootETal1992} found fluctuations in the CMB which are
of amplitude $10^{-5}$ relative to the mean background temperature of
$T_0=2.725\,\text{K}$, except for the large dipole moment, see figure
\ref{then-fig:2}, resulting in the Nobel Prize for Mather and Smoot in
2006. The small fluctuations in the CMB serve as a fingerprint of the
early universe, since the temperature fluctuations are related to the
density fluctuations at the time of last scattering. They show how
isotropic the universe was at early times. In the inflationary
scenario the fluctuations originate from quantum fluctuations which
are inflated to macroscopic scales. Due to gravitational instabilities
the fluctuations grow steadily and give rise to the formation of stars
and galaxies. \par
The theoretical framework in which the CMB and its fluctuations are
explained is Einstein's general theory of relativity. Thereby a
homogeneous and isotropic background given by a Ro\-bert\-son-Wal\-ker
universe
\cite{then:Friedmann1922,then:Friedmann1924,then:Lemaitre1927} is
perturbed. The time-evolution of the perturbations can be computed in
the framework of linear perturbation theory \cite{then:Bardeen1980}.
\par
An explanation for the presence of the CMB is the following, see also
figure \ref{then-fig:3}: We live in an expanding universe. At early
enough times the universe was so hot and dense that it was filled with
a hot plasma consisting of ionised atoms, unbounded electrons, and
photons. Due to Thomson scattering of photons with electrons, the hot
plasma was in thermal equilibrium and the mean free path of the
photons was small, hence the universe was opaque. Due to its expansion
the universe cooled down and became less dense. When the universe was
around $379\,000$ years old, its temperature $T$ has dropped down to
approximately $3000\,\text{K}$. At this time, called the time of last
scattering, the electrons got bound to the nuclei forming a gas of
neutral atoms, mainly hydrogen and helium, and the universe became
transparent. Since this time the photons travel freely on their
geodesics through the universe. At the time of last scattering the
photons had an energy distribution according to a Planck spectrum with
temperature of nearly $3000\,\text{K}$. The further expansion of the
universe redshifted the photons such that they nowadays have an energy
distribution according to a Planck spectrum with temperature of
$T_0=2.725\,\text{K}$. This is what we observe as the CMB. \par
Due to the thermal equilibrium before the time of last scattering the
CMB is nearly perfectly isotropic, but small density fluctuations lead
to small temperature fluctuations. The reason for the small
temperature fluctuations comes from a variety of effects. The most
dominant effects are the gravitational redshift that is larger in the
directions of overdense regions, a small time delay in the transition
from opaque to transparent that slightly reduces the Hubble redshift
in the directions of overdense regions, the intrinsic temperature
fluctuations, and the Doppler effect due to the velocity of the
plasma.

\section{Robertson-Walker universes} Assuming a universe whose spatial
part is locally homogeneous and isotropic, its metric is given by the
Ro\-bert\-son-Wal\-ker metric,
\begin{align*} ds^2=dt^2-\tilde{a}^2(t)\gamma_{ij}dx^{i}dx^{j},
\end{align*} where we use the Einstein summation convention. Notice
that we have changed the notation. Instead of the quaternion $z$ for
the spatial variables, we now write $x=x_0+\ii x_1+\ij x_2$.
$\gamma_{ij}$ is the metric of a homogeneous and isotropic
three-dimensional space, and the units are rescaled such that the
speed of light is $c=1$. Introducing the conformal time
$d\eta=\frac{dt}{\tilde{a}(t)}$ we have
\begin{align*}
  ds^2=a^2(\eta)\big[d\eta^2-\gamma_{ij}dx^{i}dx^{j}\big],
\end{align*} where $a(\eta)=\tilde{a}(t(\eta))$ is the cosmic scale
factor. \par
With the Ro\-bert\-son-Wal\-ker metric the Einstein equations simplify
to the Friedmann equations
\cite{then:Friedmann1922,then:Friedmann1924,then:Lemaitre1927}. One of
the two Friedmann equations reads
\begin{align*} a'^2+\kappa a^2=\frac{8\pi
    G}{3}T_{0}^{0}a^4+\frac{1}{3}\Lambda a^4 \end{align*} and the
other Friedmann equation is equivalent to local energy conservation.
$a'$ is the derivative of the cosmic scale factor with respect to the
conformal time $\eta$. $\kappa$ is the curvature parameter which we
choose to be negative, $\kappa=-1$. $G$ is Newton's gravitational
constant, $T^{\mu}_{\nu}$ is the energy-momentum tensor, and $\Lambda$
is the cosmological constant. \par
Assuming the energy and matter in the universe to be a perfect fluid
consisting of radiation, non-relativistic matter, and a cosmological
constant, the time-time component of the energy-momentum tensor reads
\begin{align*}
  T^{0}_{0}=\varepsilon_{\text{r}}(\eta)+\varepsilon_{\text{m}}(\eta),
\end{align*} where the energy densities of radiation and matter scale
like
\begin{align*}
  \varepsilon_{\text{r}}(\eta)=\varepsilon_{\text{r}}(\eta_0)\big(\frac{a(\eta_0)}{a(\eta)}\big)^4
  \quad \text{and} \quad
  \varepsilon_{\text{m}}(\eta)=\varepsilon_{\text{m}}(\eta_0)\big(\frac{a(\eta_0)}{a(\eta)}\big)^3.
\end{align*} Here $\eta_0$ denotes the conformal time at the present
epoch. \par
Specifying the initial conditions (Big Bang!) $a(0)=0,\ a'(0)>0$, the
Friedmann equation can be solved analytically
\cite{then:AurichSteiner2001},
\begin{align*}
  a(\eta)=\frac{-\big(\frac{\Omega_{\text{r}}}{\Omega_{\text{c}}}\big)^{\frac{1}{2}}{\cal
      P}'(\eta)+\frac{1}{2}\big(\frac{\Omega_{\text{m}}}{\Omega_{\text{c}}}\big)\big({\cal
      P}(\eta)-\frac{1}{12}\big)}{2\big({\cal
      P}(\eta)-\frac{1}{12}\big)^2-\frac{1}{2}\frac{\Omega_{\Lambda}\Omega_{\text{r}}}{\Omega_{\text{c}}^2}}a(\eta_0),
\end{align*} where ${\cal P}(\eta)\equiv{\cal P}(\eta;g_2,g_3)$
denotes the Weierstrass ${\cal P}$-function. The so-called invariants
$g_2$ and $g_3$ are determined by the cosmological parameters,
\begin{align*}
  g_2=\frac{\Omega_{\Lambda}\Omega_{\text{r}}}{\Omega_{\text{c}}^2}+\frac{1}{12},
  \quad
  g_3=\frac{1}{6}\frac{\Omega_{\Lambda}\Omega_{\text{r}}}{\Omega_{\text{c}}^2}-\frac{1}{16}\frac{\Omega_{\Lambda}\Omega_{\text{m}}^2}{\Omega_{\text{c}}^3}-\frac{1}{216},
\end{align*} with
\begin{align*}
  \Omega_{\text{r}}=\frac{8\pi
    G\varepsilon_{\text{r}}(\eta_0)}{3H^2(\eta_0)}, \quad
  \Omega_{\text{m}}=\frac{8\pi
    G\varepsilon_{\text{m}}(\eta_0)}{3H^2(\eta_0)}, \quad
  \Omega_{\text{c}}=\frac{1}{H^2(\eta_0)a^2(\eta_0)}, \quad
  \Omega_{\Lambda}=\frac{\Lambda}{3H^2(\eta_0)},
\end{align*}
where
\begin{align*} H(\eta)=\frac{a'(\eta)}{a^2(\eta)} \end{align*} is the
Hubble parameter. \par

\section{Perturbed Robertson-Walker universes} The idealisation to an
exact homogeneous and isotropic universe was essential to derive the
spacetime of the Ro\-bert\-son-Wal\-ker universe. But obviously, we do
not live in a universe that is perfectly homogeneous and isotropic. We
see individual stars, galaxies, and in between large empty space.
Knowing the spacetime of the Ro\-bert\-son-Wal\-ker universe, we can
study small perturbations around the homogeneous and isotropic
background. Since the amplitude of the large scale fluctuations in the
universe is of relative size $10^{-5}$, we can use linear perturbation
theory. In longitudinal gauge the most general scalar perturbation of
the Ro\-bert\-son-Wal\-ker metric reads
\begin{align*}
  ds^2=a^2(\eta)\big[(1+2\Phi)d\eta^2-(1-2\Psi)\gamma_{ij}dx^{i}dx^{j}\big],
\end{align*} where $\Phi=\Phi(\eta,x)$ and $\Psi=\Psi(\eta,x)$ are
functions of spacetime. \par
Assuming that the energy and matter density in the universe can be
described by a perfect fluid, consisting of radiation,
non-relativistic matter, and a cosmological constant, and neglecting
possible entropy perturbations, the Einstein equations reduce in first
order perturbation theory \cite{then:MukhanovFeldmanBrandenberger1992}
to
\begin{align*} &\Phi=\Psi, \\
  &\Phi''+3\hat{H}(1+c_{\text{s}}^2)\Phi'-c_{\text{s}}^2\Delta\Phi+\big(2\hat{H}'+(1+3c_{\text{s}}^2)(\hat{H}^2+1)\big)\Phi=0,
\end{align*} where $\hat{H}=\frac{a'}{a}$ and
$c_{\text{s}}^2=(3+\frac{9}{4}\frac{\varepsilon_{\text{m}}}{\varepsilon_{\text{r}}})^{-1}$
are given by the solution of the non-perturbed Ro\-bert\-son-Wal\-ker
universe. In the partial differential equation for $\Phi$ the
Laplacian occurs. If the initial and the boundary conditions of $\Phi$
are specified, the time-evolution of the metric perturbations can be
computed. \par
With the separation ansatz
\begin{align*} \Phi(\eta,x)=\sum_{k}f_k(\eta)\psi_k(x)+\int
  dk\,f_k(\eta) \psi_k(x), \end{align*} where the $\psi_k$ are the
eigenfunctions of the negative Laplacian, and the $E_k$ are the
corresponding eigenvalues,
\begin{align*} -\Delta\psi_k(x)=E_k\psi_k(x), \end{align*} the partial
differential equation for $\Phi$ simplifies to
\begin{align*}
  f_k''(\eta)+3\hat{H}(1+c_{\text{s}}^2)f_k'(\eta)+\big(c_{\text{s}}^2E_k+2\hat{H}'+(1+3c_{\text{s}}^2)(\hat{H}^2+1)\big)f_k(\eta)=0.
\end{align*} These ordinary differential equations (one for each
eigenvalue $E_k$) can be computed numerically in a straightforward
way, and we finally obtain the metric of the whole universe. This
gives the input to the Sachs-Wolfe formula which connects the metric
perturbations with the temperature fluctuations,
\begin{align*} \frac{\delta
    T}{T_0}(\hat{n})=2\Phi(\eta_{\text{SLS}},x(\eta_{\text{SLS}}))-\frac{3}{2}\Phi(0,x(0))+2\int_{\eta_{\text{SLS}}}^{\eta_0}d\eta\,\frac{\partial}{\partial\eta}\Phi(\eta,x(\eta)),
\end{align*} where $\hat{n}$ is a unit vector in the direction of the
observed photons. $x(\eta)$ is the geodesic along which the light
travels from the surface of last scattering (SLS) towards us, and
$\eta_{\text{SLS}}$ is the time of last scattering. \par
If we choose the topology of the universe to be the Picard surface, we
can use the Maass waveforms of sections \ref{then-sec:5} and
\ref{then-sec:6} in the separation ansatz for the metric perturbations
$\Phi$. Let us further choose the initial conditions to be
\begin{align*} f_k(0)=\frac{\sigma_k\alpha}{\sqrt{E_k\sqrt{E_k-1}}}
  \quad \text{and} \quad
  f_k'(0)=\frac{-\Omega_{\text{m}}f_k(0)}{16(\Omega_{\text{c}}\Omega_{\text{r}})^{\frac{1}{2}}},
\end{align*} \cite{then:AurichSteiner2001}, which carry over to a
Har\-ri\-son-Zel'\-do\-vich spectrum. $\alpha$ is a constant
independent of $k$ which is fitted to the amplitude of the observed
temperature fluctuations. The quantities $\sigma_k$ are normal
distributed random numbers. \par
\begin{figure} \centering
  \includegraphics[height=11cm,angle=-90]{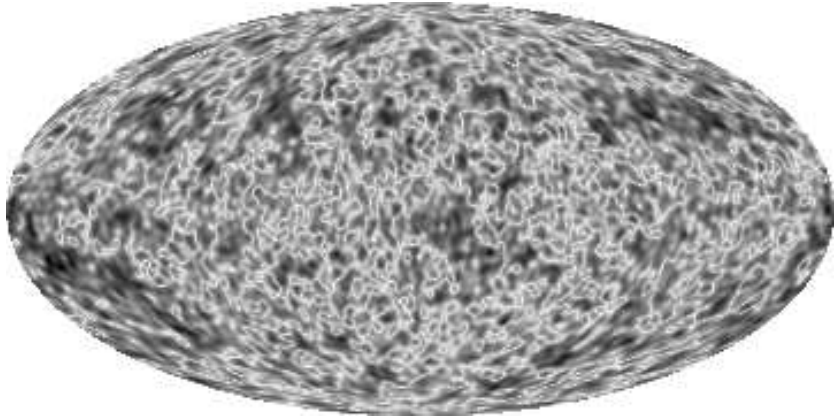}
  \caption{The sky map of the calculated temperature fluctuations of
    the CMB for
    $\Omega_{\text{tot}}=0.95,\Omega_{\text{m}}=0.3,\Omega_{\Lambda}=0.65,h_0=0.65$,
    and $x_{\text{obs}}=0.2+0.1\ii+1.6\ij$, if the discrete spectrum
    with $E_k=k^2+1\le19601$ is taken into account. (The figure is
    taken from \cite{then:AurichLustigSteinerThen2004}).}
  \label{then-fig:4} \end{figure}

The following cosmological parameters are used
$\Omega_{\text{m}}=0.3,\Omega_{\Lambda}=0.65,\Omega_{\text{c}}=1-\Omega_{\text{tot}}=1-\Omega_{\text{r}}-\Omega_{\text{m}}-\Omega_{\Lambda},H(\eta_0)=100\,h_0\,\text{km}\,\text{s}^{-1}\,\text{Mpc}^{-1}$
with $h_0=0.65$. The density $\Omega_{\text{r}}\approx10^{-4}$ is
determined by the current temperature $T_0=2.725\,\text{K}$. The point
of the observer is chosen to be at $x_{\text{obs}}=0.2+0.1\ii+1.6\ij$.
For numerical reasons the infinite spectrum is cut such that only the
discrete eigenvalues with $E_k=k^2+1\le19601$ and their corresponding
eigenfunctions are taken into account. The necessary computations are
carried out in \cite{then:AurichLustigSteinerThen2004}. The resulting
sky map is shown in figure \ref{then-fig:4}.

\begin{figure} \centering
  {\scriptsize$C(\vartheta)\ [\mu\text{K}^2]$} \hspace*{160pt} \\
  \includegraphics[width=200pt]{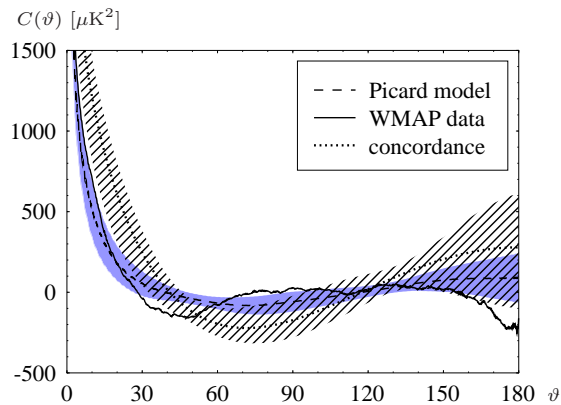} {\scriptsize$\vartheta$}
  \caption{Correlation functions of the calculated temperature
    fluctuations on the Picard surface (dashed line)
    \cite{then:AurichLustigSteinerThen2005}, of the WMAP observations
    (solid line) \cite{then:SpergelETal2003}, and of the concordance
    model (dotted line) \cite{then:SpergelETal2003}. (The figure is
    borrowed from \cite{then:AurichLustigSteinerThen2005}).}
  \label{then-fig:5} \end{figure}

Concerning the topology of the universe which manifests itself in the
suppression of power in the large scale anisotropies, there exist the
cosmological observations from COBE and WMAP. In order to
quantitatively compare our results with these observations, we
introduce the two-point correlation function,
\begin{align*} C(\vartheta)=\big\langle\delta T(\hat{n})\delta
  T(\hat{n}')\big\rangle_{\cos\vartheta=\hat{n}\cdot\hat{n}'}.
\end{align*} Figure \ref{then-fig:5} shows the correlation function
corresponding to the calculated sky map of the Picard surface
\cite{then:AurichLustigSteinerThen2005} in comparison with the results
of the cosmological observations \cite{then:SpergelETal2003} and with
the concordance model \cite{then:SpergelETal2003}.

We find quite good agreement of our calculated temperature
fluctuations with the cosmological observations, whereas the results
of the concordance model is not in good agreement with the data for
$\vartheta\gtrsim7^{\circ}$. Especially for large angular separations,
$\vartheta\gtrsim160^{\circ}$, the concordance model is not able to
describe the observed anticorrelation in $C(\vartheta)$. This
anticorrelation constitutes a fingerprint in the CMB that favours a
non-trivial topology for the universe.

\section{Acknowledgments}
The author thanks Professor Hishamuddin Zainuddin for the invitation
to the Theoretical Studies Laboratory at the Universiti Putra Malaysia
and for the pleasant stay there. Collaborations with Ralf Aurich, Sven
Lustig, and Frank Steiner are gratefully acknowledged. Highest thanks
are due to Dennis A.~Hejhal for sharing his knowledge with me.  Part
of the work has been supported by the European Commission under the
Research Training Network (Mathematical Aspects of Quantum Chaos) no
HPRN-CT-2000-00103.  The free access to the Legacy Archive for
Microwave Background Data Analysis (LAMBDA) \cite{then:LAMBDA2003} is
appreciated. Support for LAMBDA is provided by the NASA Office of
Space Science.  The computations were run on the computers of the
Universit\"{a}ts-Rechenzentrum Ulm.

\end{document}